# Inherited Weak Topological Insulator Signatures in Topological Hourglass Semimetal $Nb_3XTe_6$ (X = Si, Ge)

Q. Wan[1,#], T. Y. Yang[1,#], S. Li[2], M. Yang[3,4], Z. Zhu[5], C. L. Wu[1], C. Peng[1], S. K. Mo[1], W. K. Wu[2], Z. H. Chen[6], Y. B. Huang[6], L. L. Lev[7,8], V. N. Strocov[7], J. Hu[9], Z. Q. Mao[10], Hao Zheng[5], J. F. Jia[5], Y. G. Shi[3,4,11], Shengyuan A. Yang[2] and N. Xu[1,*]

[1] *Institute of Advanced Studies, Wuhan University, Wuhan 430072, China*
[2] *Research Laboratory for Quantum Materials, Singapore University of Technology and Design, Singapore 487372, Singapore*
[3] *Beijing National Laboratory for Condensed Matter Physics and Institute of Physics, Chinese Academy of Sciences, Beijing 100190, China*
[4] *School of Physical Sciences, University of Chinese Academy of Sciences, Beijing 100190, China*
[5] *Key Laboratory of Artificial Structures and Quantum Control (Ministry of Education), Shenyang National Laboratory for Materials Science, School of Physics and Astronomy, Shanghai Jiao Tong University, 800 Dongchuan Road, Shanghai 200240, China*
[6] *Shanghai Synchrotron Radiation Facility, Shanghai Advanced Research Institute, Chinese Academy of Sciences, Shanghai 201204, China*
[7] *Swiss Light Source, Paul Scherrer Institut, CH-5232 Villigen PSI, Switzerland*
[8] *Moscow Institute of Physics and Technology, 9 Institutskiy per., Dolgoprudny, Moscow Region, 141701, Russia*
[9] *Department of Physics, University of Arkansas, Fayetteville, AR 72701, USA.*
[10] *Department of Physics, Pennsylvania State University, University Park, PA 16803, USA.*
[11] *Songshan Lake Materials Laboratory, Dongguan, Guangdong 523808, China*

*# These authors contributed equally to this work.*
*\* E-mail: nxu@whu.edu.cn*

# Abstract

Using spin-resolved and angle-resolved photoemission spectroscopy and first-principles calculations, we have identified bulk band inversion and spin polarized surface state evolved from a weak topological insulator (TI) phase in van der Waals materials $Nb_3XTe_6$ (X = Si, Ge). The fingerprints of weak TI homologically emerge with hourglass fermions, as multi nodal chains composed by the same pair of valence and conduction bands gapped by spin orbit coupling. The novel topological state, with a pair of valence and conduction bands encoding both weak TI and hourglass semimetal nature, is essential and guaranteed by nonsymmorphic symmetry. It is distinct from TIs studied previously based on band inversions without symmetry protections.

## I. INTRODUCTION

The topological property of wave functions is a key characterization of quantum materials [1-5]. Novel boundary states provide an observable signature of the nontrivial topological invariant of wave functions, and topological phase transitions accompany with modifications of boundary states. From band theory point of view, band inversion (Fig. 1a-b) plays a critical role in formation of a topological state. In a topological insulator (TI) [6-16], spin-orbit coupling (SOC) opens a bulk hybridization gap between the inverted bands, and Dirac surface states emerge in the gap with spin-momentum locking (Fig. 1c). Benefited from specific crystalline symmetries, nodes can survive within the bulk hybridization gap (Fig. 1d) and define topological semimetals (TSM) [17-37]. Tuning the hopping term, SOC, onsite and long-range Coulomb interactions *etc.*, topological phase transitions can be realized by removing the band inversion without changing symmetry [38-40], as described in Fig. 1b. However, wave function simultaneously embedding multiple types of topological phases is elusive and the properties of such state is unknown.

Here, we unveil a novel topological phase hosting both weak TI and hourglass TSM fingerprints in van der Waals (vdW) materials $Nb_3XTe_6$ (X = Si, Ge). It is evolved from SOC gapped multi nodal chains composed by the same pair of valence and conduction bands. Combining bulk-sensitive soft-x-ray angle-resolved photoemission spectroscopy (SX-ARPES), spin-resolved ARPES with surface-sensitive vacuum ultraviolet (VUV) light and first-principles calculations, we uncover bulk band inversion and spin polarized topological surface states induced by a weak TI phase with a unique topological invariant $Z_2 = \{0;110\}$. The weak TI signatures are inherited as hourglass nodes close the hybridization gap as imposed by nonsymmorphic symmetry. We further demonstrate that topological phase in $Nb_3XTe_6$ is based on a band inversion guaranteed by nonsymmorphic symmetry, which, in contrast to TI based on band inversion without symmetry protection as depicted in Fig. 1a-c, is essential and cannot be removed without breaking the nonsymmorphic symmetry. Our results not only realize a novel topological phase hosting distinctions of both weak TI and TSM phases by the same group of valence and conduction bands, but also provide a striking vdW material platform for fine tuning the topological states, bulk/surface hourglass fermions and their interplay with multiple parameters including thickness and Si/Ge stoichiometry.

$Nb_3XTe_6$ belongs to a vdW material family $Nb_{2n+1}XTe_{4n+2}$, with crystal structure and Brillouin zones (BZ) shown in Fig. 1e-f, respectively. In addition to the time reversal symmetry $T$ and inversion symmetry $P$, there are two glide mirror planes perpendicular to the x and y directions,

$$\widetilde{M_x}(x,y,z) \to (-x+1/2, y+1/2, z+1/2) \text{ and } \widetilde{M_y}(x,y,z) \to (x+1/2, -y+1/2, z),$$

respectively, characterizing the nonsymmorphic feature of the system. SOC strength in $Nb_3XTe_6$ is considerably strong due to the heavy Nb atoms. Previous theoretical study [41] predicts bulk hourglass fermions, with transport and ARPES evidences reported in a sister compound $Ta_3SiTe_6$ [42-44]. The system exhibits multi parameters for controlling the electronic and topological properties. Few layers $Nb_3SiTe_6$ flakes have been successfully fabricated and enhanced electron coherence was reported [45]. By fine tuning stoichiometry, directional massless Dirac fermions have been achieved in $NbSi_{0.45}Te_2$ with one-dimensional (1D) long range order [46].

## II. EXPERIMENTAL AND COMPUTATIONAL DETAILS

The $Nb_3SiTe_6$ single crystals were synthesized with a mixture of Nb, Si and Te at a molar ratio of 3:1:6 using chemical vapour transport. During growth of $Nb_3SiTe_6$ the temperature was set at 950 °C and 850 °C, respectively, for the hot and cold ends of the double zone tube furnace. Single crystals of $Nb_3GeTe_6$ were grown by a solid-state reaction method. A mixture of Nb powder, Ge piece and Te granulum at a molar ratio of 3:1:6 was pulverized, pressed into a pellet, placed in an alumina crucible, and then sealed in a highly evacuated quartz tube. After that, the tube was heated to 1100 °C for 10 hours and slowly cooled down to 650 °C at a rate of 2 °C/hour.

Clean surfaces for ARPES measurements were obtained by cleaving samples *in situ* in a vacuum better than $5 \times 10^{-11}$ Torr. VUV and SX-ARPES experiments were performed at Dreamline of Shanghai Synchrotron Radiation Facility and SX-ARPES endstation of the ADRESS beamline at the Swiss Light Source, Paul Scherrer Institute, Switzerland [47], respectively. Spin-resolved ARPES measurements were done with a home-designed ARPES facility equipped a VLEED spin detector, with a Sherman function of S = 0.27 used to generate the measured spin polarizations.

First-principles calculations were performed within the framework of density functional theory using the Perdew-Burke-Ernzerhof-type [48] generalized gradient approximation for the exchange correlation functional as implemented in the Vienna ab initio simulation package [49-51]. The BZ was sampled with Γ–centered k mesh of size 8×5×4, and the cutoff energy was set as 350 eV. The energy and force convergence criteria were set to be $10^{-5}$ eV and 0.01 eV/Å, respectively. The van der Waals (vdW) corrections have been taken into account with the approach of Dion et al. [52]. The surface states were calculated by constructing the maximally localized Wannier functions [53, 54] and by using the iterative Green function method [55, 56] as implemented in the WannierTools package [57].

### III. RESULTS

Figure 1g shows the ARPES spectra of $Nb_3SiTe_6$ acquired with $h\nu$ = 300~570 eV. A $k_z$ dispersive feature confirms the bulk origin of the photoelectron intensity in SX-ARPES measurements, due to the increase of the photoelectron mean free path compared to the VUV energy range [58]. A heavy band (named as α) and a light band (named as β) are resolved along the Γ-Z direction. Compared to calculations, they are visible in every other BZ due to a photoemission selection rule [59-61] related to nonsymmorphic symmetry $\widetilde{M_x}$ in $Nb_3SiTe_6$. Similarly, we observe single branch of Dirac-cone-like dispersion along the Z-S direction in Fig. 1h due to $\widetilde{M_y}$. Off of the photoemission scattering plan, the selection rule is not strictly applied and both branches of Dirac-cone-like dispersions are visible along the T-R direction. Our results suggest a quasi-1D Dirac-cone-like dispersion along the S-R direction, consistent with the nodal-line along the S-R high symmetry line (NL$_{SR}$) protected by the interplay of $\widetilde{M}_x$, $T\widetilde{M}_y$ and $PT$, as indicated by the calculations without SOC (black line in Fig. 1h).

Furthermore, we observe multi band inversions between the α and β bands, forming two kinds of nodal chains with finite and infinite length, respectively, if ignoring SOC. Along the Γ-Y direction (Fig. 1i-j), besides the α and β bands in the first BZ (red marks in Fig. 1j), the α' and β' bands are expected in the 2$^{nd}$ BZ (blue marks in Fig. 1j) with zero spectra weight due to selection rule related to $\widetilde{M}_x$. Because both the α and β bands cross E$_F$ along the Γ-Y direction, the bottom/top of the α/β band locates below/above E$_F$ at the Γ/Γ' point. Therefore, the α and β' bands have to overlap with each other at the Γ point as reproduced by first-principles calculations (Fig. 1j). The α band bottom is ~100 meV below E$_F$, in the same order of energy resolution of SX-ARPES measurements. Although we cannot clearly resolve the parabolic dispersion within this small energy window, the DFT calculations show an overall agreement with SX-ARPES results (Fig. 1i-j). The band inversion between the α and β' bands around the Γ point leads to a nodal ring in the $k_x = 0$ plane (labeled as NR$_x$), which is protected by $\widetilde{M}_x$ when SOC is not included (Supplementary Materials Fig. S1-2).

Along the Γ-X direction, the α' and β' bands can be resolved in the 2$^{nd}$ BZ, which are quite close to each other near E$_F$ and well separated at E$_B$ > 0.2 eV. Similarly, because both the α' and β' bands cross E$_F$, we can ascertain that the β' band crosses with the α band with missing spectra weight near the X point, forming a nodal chain touching at the S points in the $k_y = 0$ plane (labeled as NC$_y$), which is guaranteed by nonsymmorphic symmetries as discussed in details later.

Figure 2a summarizes the nodes composed by the α/α' and β/β' bands when SOC is not included. First-principles calculations further unveil two more nodal rings touching with NR$_x$ and forming a nodal chain with a finite length (NC$_x$). Topological drumhead surface states are shown in the calculated spectra function on the (001) surface (Fig. 2b), with the boundary bonding to surface projections of the nodes. The drumhead states are four-fold degenerated (including the spin degeneracy), because each point on the nodal chain projections line corresponds to two nodes with ±k$_z$ in the 3D BZ. This is consistent with previous theoretical study on topological nodal chain semimetal IrF$_4$ [62].

In the presence of SOC, the nodal line and chains are gapped, however, in two distinct ways. SOC gaps NL$_{SR}$ as described in Fig. 1d, with hourglass nodes forming rings NR$_{HG}$ surrounding the S points in Fig. 2c (Supplementary Materials Fig. S3-4). For NC$_x$ and NC$_y$, SOC fully gaps the nodes and results in a TI scenario as described in Fig. 1c. Parity calculations indicate Z$_2$ indices of {0;110} in a weak TI configuration with band inversion happened at the Γ and U points (Fig. 2c). Topological surface states related to the weak TI phase emerge on the (001) surface (Fig. 2d). Compare to the calculation without SOC, the bulk band projections open a gap at the $\bar{\Gamma}$ point, and a pair of Dirac cones appears inside the gap with Dirac points close to the bulk valence and conduction bands, respectively (Fig. 2e). We note that $\widetilde{M_y}$ is preserved on the $\bar{\Gamma}$-$\bar{X}$ path near (001) surface region, therefore along the $\bar{\Gamma}$-$\bar{X}$ direction such a dual-Dirac-cone structure is an effective hourglass dispersion as imposed by $\widetilde{M_y}$.

The spin polarized topological surface states is directly observed in our spin resolved VUV-ARPES measurements on Nb$_3$GeTe$_6$, which shares a similar electronic structure with Nb$_3$SiTe$_6$ (Supplementary Materials Fig. S5). In Fig. 3a, we clearly identify a band near the $\bar{X}$ point showing no dispersion with photon energy in the range of 24~90 eV, which conforms the 2D surface state nature. We plot the FS mapped in the surface BZ (Fig. 3b). Figures 3c and d display the surface band structure along the $\bar{\Gamma}$-$\bar{X}$ and $\bar{Y}$-$\bar{U}$ directions, respectively, in which additional spectra weight appears near E$_F$ around the $\bar{X}$ and $\bar{U}$ points, in a good agreement with the topological surface state indicated by calculations in Fig. 2d.

We performed spin-resolved ARPES measurements to study the spin texture of the surface states. We note that the upper branch of lower Dirac cone (DC$_l$) and lower branch of upper Dirac cone (DC$_U$) are quite close to each other near the $\bar{X}$ point (Fig. 2d) and the spin signals would be mixed by limited experimental resolution. To avoid this complexity, we focus on the surface states along the $\bar{\Gamma}$-$\bar{X}$ direction in which only single Dirac cone dispersion has non-zero spectra weight in ARPES experiment, due to the selection rule related to $\widetilde{M_y}$. Figures 3e and f show the spin-resolved EDC intensity I$^{\uparrow\downarrow}_{x,y}$ in the x and y directions, respectively, measured at E point labeled in Fig. 3b and c. While I$^{\uparrow}_x$ is almost equal to I$^{\downarrow}_x$ (Fig. 3e), there is a clear difference in I$^{\uparrow}_y$ and I$^{\downarrow}_y$ at the EDC's peak that corresponds to surface states (Fig. 3f), indicating that the observed surface state is spin-polarized along the y direction. At the time reversal symmetric F point labeled in Fig. 3b and c, the surface state is spin-polarized along the y direction (Fig. 3g), however, in the opposite direction to that at the E point. Our spin-resolved

VUV-ARPES clearly reveal the topological surface states with spin-momentum locking as evidence of inherited weak TI phase in Nb$_3$XTe$_6$.

The bulk bands of Nb$_3$XTe$_6$ are spin-degenerated as required by time reversal and inversion symmetries. The "hidden" spin signals from the bulk projection bands, as observed in systems with both time reversal and inversion symmetries [63-64], are forbidden along the $\bar{\Gamma}$ - $\bar{X}$ direction by symmetries of Nb$_3$XTe$_6$ monolayer (Supplementary Materials Fig. S6). Therefore, spin polarization observed in Fig. 3e-g are from topological surface states, without contribution from bulk band projections.

## IV. DISCUSSIONS

Figure 4 schematically summarizes our main finding of the electronic structure and topological properties of Nb$_3$XTe$_6$. From SX-ARPES results in Fig. 1, we directly observe the bulk α and β bands overlap with each other near E$_F$ and form nodal line NL$_{SR}$ and nodal chains NC$_x$/NC$_y$ if ignoring SOC (Fig. 4a). Topological drumhead state emerges on the (001) surface, bonding to the surface projections of nodes (Fig. 4b). In contrast to the nodal line/chain induced by band overlap without symmetry protection as depicted in Fig. 1a-b, NC$_y$ is essential and based on a band inversion (Fig. 1k) guaranteed by nonsymmorphic symmetries, as illustrated by Fig. 4c-d. Because the k$_y$ = 0 plane, which hosts the nodal chain NC$_y$ (yellow plane in Fig. 4a and c), is the invariant plane of $\widetilde{M_y}$, for each band $|u\rangle$ we have

$$\widetilde{M_y}|u\rangle = g_{\widetilde{M_y}}|u\rangle.$$

Because of non-primitive translation operations, $\widetilde{M_y}$ takes eigenvalues of

$$g_{\widetilde{M_y}} = \pm e^{-\frac{ik_x a}{2}},$$

when SOC is not included. We now consider band structure along the path of K-Γ-Q in Fig. 4c, where K = (π, 0, k$_z$) and Q = (k$_x$, 0, π) are arbitrary points sitting on the X-S and S-Z high symmetry lines, respectively. The solid and dashed lines in Fig. 4d distinguish the additional phase term (-1) in $g_{\widetilde{M_y}}$. Because the K point on the X-S high symmetry line is invariant under $\widetilde{M_x}$, and the commutation relationship between $\widetilde{M_x}$ and $\widetilde{M_y}$ is

$$\widetilde{M_y}\widetilde{M_x} = T_{1\bar{1}0}\widetilde{M_x}\widetilde{M_y},$$

where $T_{1\bar{1}0}(x, y, z) = (x + a, y - b, z)$,

the bands with additional phase term have to degenerate at the K point with the opposite of eigenvalues of $\pm i$ in Fig. 4d:

$$\widetilde{M_y}\widetilde{M_x}|u\rangle_K = e^{-ik_x a}\widetilde{M_x}\widetilde{M_y}|u\rangle_K = -g_{\widetilde{M_y}}\widetilde{M_x}|u\rangle_K.$$

Because there are two layers in the unit cell of Nb$_3$XTe$_6$ (Fig. 1e) and the interlayer coupling strength is considerable (Fig. 1g), the interlayer splitting has to be taken into accounted in Fig. 4d. The double degeneracy on the S-Z high-symmetry line is guaranteed by $T\tilde{C}_{2z}$ symmetry, where $\tilde{C}_{2z}$ is a screw rotation along z direction, i.e.

$$\tilde{C}_{2z}(x,y,z) \rightarrow (-x,-y,z+1/2c).$$

At $k_z = \pi$ plane which includes the S-Z path and is invariant under $T\tilde{C}_{2z}$, $(T\tilde{C}_{2z})^2 = -1$ ensures each band to be Kramers doubly degenerate. Also, by noting

$$\widetilde{M_y}(T\tilde{C}_{2z}) = T_{110}(T\tilde{C}_{2z})\widetilde{M_y} = e^{-i(k_x a + k_y b)}(T\tilde{C}_{2z})\widetilde{M_y} = e^{-i(k_x a)}(T\tilde{C}_{2z})\widetilde{M_y},$$

along the S-Z line, each Kramers pair at the Q point must share the same $\widetilde{M_y}$ eigenvalue, as imposed by

$$\widetilde{M_y}(T\tilde{C}_{2z})|u\rangle_Q = e^{-i(k_x a)}(T\tilde{C}_{2z})\widetilde{M_y}|u\rangle_Q = g_{\widetilde{M_y}}(T\tilde{C}_{2z})|u\rangle_Q,$$

$$\text{where } g_{\widetilde{M_y}} = e^{-\frac{ik_x a}{2}} \text{ or } -e^{-\frac{ik_x a}{2}}.$$

Thus, the band dispersion along the K-Γ-Q path effectively forms an hourglass dispersion, with the hourglass node with opposite $g_{\widetilde{M_y}}$ values guaranteed by nonsymmorphic symmetries. It is different from band inversion without symmetry protection as described in Fig. 1b. By tuning hopping parameters, the band inversion along the Γ-K direction can be lifted and the hourglass nodes will be shifted to the Γ-Q direction but cannot be eliminated without breaking the nonsymmorphic symmetries (Supplementary Materials Fig. S7).

The considerable strength of SOC in Nb$_3$XTe$_6$ eventually gaps the nodal line and chains in Fig. 4a in two ways. As protected by $\widetilde{M_x}$, SOC turns NL$_{SR}$ into an hourglass nodal ring NR$_{HG}$ (Fig. 4e). In contrast, SOC fully gaps NC$_x$ and NC$_y$ and results in a weak TI phase with Z$_2$ = {0;110}. Topological drumhead surface states evolve into a pair of spin split Dirac cones at the $\bar{\Gamma}$ point in Fig. 4f. Such a dual-Dirac-cone structure is effectively an hourglass dispersion along $\bar{\Gamma}$-$\bar{X}$ direction, because the $\widetilde{M_y}$ is preserved on the $\bar{\Gamma}$-$\bar{X}$ path. The upper branches of the two Dirac cones have to be degenerated at the $\bar{X}$ point as required by time reversal symmetry, through the bulk conduction bands. Similarly, the lower branches of Dirac cones have to be degenerated at the $\bar{X}$ point through the bulk valence bands. An hourglass node is formed and guaranteed by $\widetilde{M_y}$ (inset of Fig. 4f). The spin polarized topological surface state, as the signature of weak TI phase, are observed in spin-resolved VUV-ARPES measurements in Fig. 3, coexisting with bulk hourglass fermions.

Therefore, we uncover a novel topological phase in $Nb_3XTe_6$, which shows full set of fingerprints of weak TI and hourglass TSM phases near $E_F$, derived from the same group of valence and conduction bands in single material. The inherited TI signatures are stable and guaranteed by nonsymmorphic symmetries, in contrast to that induced by band inversion without symmetry protection. The novel behaviors of topological hourglass surface state on the (001) surface with a narrow band-width inherited from weak TI phase and the interactions with homological bulk hourglass fermions call for further research activities. Considering the compositional-tunable nature of the layered vdW $NbX_nTe_2$ family [65] and 1D massless Dirac fermions observed in $NbSi_{0.45}Te_2$ [46], our results also unveil a promising platform for tuning the hourglass fermions and topological phase transition with multiple parameters such as thickness and X stoichiometry.

## V. ACKNOWLEDGEMENT

This work was supported by the Ministry of Science and Technology of China (grant no. 2018YFA0307000), Chinese National Key Research and Development Program (grant nos. 2018FYA0305800, 2017YFA0302901, 2016YFA0300604), the National Natural Science Foundation of China (U2032128, U2032204, 11874047, 92065201, 11790313), Singapore Ministry of Education AcRF Tier 2 (Grants No. MOE2019-T2-1-001), the Strategic Priority Research Program (B) of the Chinese Academy of Sciences (No. XDB33000000).

## References

[1] K. v. Klitzing, G. Dorda and M. Pepper, *New method for High-Accuracy Determination of the Fine-Structure Constant Based on Quantized Hall Resistance.* Phys. Rev. Lett. **45**, 494 (1980).

[2] D. J. Thouless, M. Kohmoto, M. P. Nightingale and M. den Nijs, *Quantized Hall Conductance in a Two-Dimensional Periodic Potential.* Phys. Rev. Lett. **49**, 405 (1982).

[3] M. Kohmoto, T*opological invariant and the quantization of the Hall conductance.* *Ann. Phys.* **160**, 343 (1985).

[4] X. G. Wen, *Topological Orders in Rigid States.* Int. J. Mod. Phys. B **4**, 239 (1990).

[5] X. G. Wen, *Theory of The Edge States in Fractional Quantum Hall Effects.* Int. J. Mod. Phys. B **6**, 1711 (1992).

[6] M. Z. Hasan and C. L. Kane, *Colloquium: Topological insulators*. Rev. Mod. Phys. **82**, 3045 (2010).

[7] X. L. Qi and S. C. Zhang, *Topological insulators and superconductors*. Rev. Mod. Phys. **83**, 1057 (2011).

[8] J. E. Moore and L. Balents, *Topological invariants of time-reversal-invariant band structures*. Phys. Rev. B **75**, 121306 (2007).

[9] C. L. Kane and E. J. Mele, *Z2 Topological Order and the Quantum Spin Hall Effect.* Phys. Rev. Lett. **95**, 146802 (2005).

[10] C. L. Kane and E. J. Mele, *Quantum Spin Hall Effect in Graphene*. Phys. Rev. Lett. **95**, 226801 (2005).

[11] B. A. Bernevig, T. L. Hughes and S. C. Zhang, *Quantum Spin Hall Effect and Topological Phase Transition in HgTe Quantum Wells*, Science **314**, 1757 (2006).

[12] M. König, S. Wiedmann, C. Brüne, A. Roth, H. Buhmann, L. W. Molenkamp, X. L. Qi and S. C. Zhang, *Quantum Spin Hall Insulator State in HgTe Quantum Wells*. Science **318**, 766 (2007).

[13] L. Fu and C. L. Kane, *Topological Insulators with Inversion Symmetry*. Phys. Rev. B **76,** 045302 (2007).

[14] C. L. Kane and E. J. Mele, *A New Spin on the Insulating State.* Science **314**, 1692 (2006).

[15] H. J. Zhang, C. X. Liu, X. L. Qi, X. Dai, Z. Fang and S. C. Zhang, *Topological insulators in Bi2Se3, Bi2Te3 and Sb2Te3 with a single Dirac cone on the surface.* Nat. Phys. **5** 438 (2009).

[16] Y. L. Chen*,* J. G. Analytis, J. H. Chu, Z. K. Liu, S. K. Mo, X. L. Qi, H. J. Zhang, D. H. Lu, X. Dai, Z. Fang, S. C. Zhang, I. R. Fisher, Z. Hussain and Z. X. Shen, *Experimental Realization of a Three-Dimensional Topological Insulator, $Bi_2Te_3$*, Science **325**, 178 (2009).

[17] N. P. Armitage, E. J. Mele, and Ashvin Vishwanath, *Weyl and Dirac semimetals in three-dimensional solids*, Rev. Mod. Phys. 90, 015001 (2018).

[18] Z. J. Wang, Y. Sun, X. Q. Chen, C. Franchini, G. Xu, H. Weng, X. Dai, and Z. Fang, *Dirac semimetal and topological phase transitions in $A_3Bi$ (A = Na, K, Rb)*. Phys. Rev. B **85**, 195320 (2012).

[19] Z. K. Liu, B. Zhou, Y. Zhang, Z. J. Wang, H. M. Weng, D. Prabhakaran, S.-K. Mo, Z. X. Shen, Z. Fang, X. Dai, Z. Hussain, and Y. L. Chen, *Discovery of a Three-Dimensional Topological Dirac Semimetal, $Na_3Bi$*. Science **343**, 864 (2014).

[20] M. Neupane, S. Y. Xu, R. Sankar, N. Alidoust, G. Bian, C. Liu, I. Belopolski, T. R. Chang, H. T. Jeng, H. Lin, A. Bansil, F. Chou and M. Z. Hasan, *Observation of a three-dimensional topological Dirac semimetal phase in high-mobility Cd3As2*. Nat. Commun. **5**, 3786 (2014).

[21] S. Borisenko, Q. Gibson, D. Evtushinsky, V. Zabolotnyy, B. Buchner and R. J. Cava, *Experimental Realization of a Three-Dimensional Dirac Semimetal*. Phys. Rev. Lett. **113**, 027603 (2014).

[22] J. Xiong, S. K. Kushwaha, T. Liang, J. W. Krizan, M. Hirschberger, W. Wang, R. J. Cava, and N. P. Ong, *Evidence for the chiral anomaly in the Dirac semimetal $Na_3Bi$*. Science **350**, 413 (2015).

[23] L. M. Schoop, M. N. Ali, C Straer, A. Topp, A. Varykhalov, D. Marchenko, V. Duppel, S. S. P. Parkin, B. V. Lotsch and C. R. Ast, *Dirac cone protected by non-symmorphic symmetry and three-dimensional Dirac line node in ZrSiS*. Nat. Commun. **7**, 11696 (2016).

[24] H. M. Weng, Y. Y. Liang, Q. N. Xu, R. Yu, Z. Fang, X. Dai and Y. Kawazoe, *Topological node-line semimetal in three-dimensional graphene networks*. Phys. Rev. B **92**, 045108 (2015).

[25] R. Yu, H. M. Weng, Z. Fang, X Dai and X. Hu, *Topological Node-Line Semimetal and Dirac Semimetal State in Antiperovskite $Cu_3PdN$*. Phys. Rev. Lett. **115**, 036807 (2015).

[26] G. Bian, T.-R. Chang, R Sankar, S.-Y. Xu, H. Zheng, T. Neupert, C-K Chiu, S.-M. Huang, G. Q. Chang, I. Belopolski1, D. S. Sanchez, M. Neupane, N. Alidoust, C. Liu, B. K. Wang, C.-C. Lee, H.-T. Jeng, C. L. Zhang, Z. J. Yuan, S. Jia, A. Bansil, F. C. Chou, H. Lin and M. Z. Hasan, *Topological nodal-line fermions in spin-orbit metal $PbTaSe_2$*. Nat. Commun. **7**, 10556 (2016).

[27] J. Hu, Z. J. Tang, J. Y. Liu, X. Liu, Y. L. Zhu, D. Graf, K. Myhro, S. Tran, C. N. Lau, J. Wei and Z. Q. Mao, *Evidence of topological nodal-line fermions in ZrSiSe*

*and ZrSiTe.* Phys. Rev. Lett. **117**, 016602 (2016).

[28] X. Wan, A. M. Turner, A. Vishwanath and S. Y. Savrasov, *Topological semimetal and Fermi-arc surface states in the electronic structure of pyrochlore iridates.* Phys. Rev. B **83**, 205101 (2011).

[29] G. Xu, H. Weng, Z. Wang, X. Dai and Z. Fang, *Chern Semimetal and the Quantized Anomalous Hall Effect in $HgCr_2Se_4$.* Phys. Rev. Lett. **107**, 186806 (2011).

[30] H. M. Weng, C. Fang, Z. Fang, B. A. Bernevig and X. Dai, *Weyl Semimetal Phase in Noncentrosymmetric Transition-Metal Monophosphides.* Phys. Rev. X **5**, 011029 (2015).

[31] S.-M. Huang, S.-Y. Xu, I. Belopolski, C.-C. L, G. Q. Chang, B. K. Wang, N. Alidoust, G. Bian, M. Neupane, C. L. Zhang, S. Jia, A. Bansil, H. Lin and M. Z. Hasan, *A Weyl Fermion semimetal with surface Fermi arcs in the transition metal monopnictide TaAs class.* Nat. Commun. **6**, 7373 (2015).

[32] B. Q. Lv, H. M. Weng, B. B. Fu, X. P. Wang, H. Miao, J. Ma, P. Richard, X. C. Huang, L. X. Zhao, G. F. Chen, Z. Fang, X. Dai, T. Qian and H. Ding, *Experimental Discovery of Weyl Semimetal TaAs.* Phys. Rev. X **5**, 031013 (2015).

[33] B. Q. Lv, N. Xu, H. M. Weng, J. Z. Ma, P. Richard, X. C. Huang, L. X. Zhao, G. F. Chen, C. Matt, F. Bisti, V. N. Strocov, J. Mesot, Z. Fang, X. Dai, T. Qian, M. Shi and H. Ding, Nat. Phys. **11**, 724 (2015).

[34] S. Y. Xu, I. Belopolski, N. Alidoust, M. Neupane, G. Bian, C. L Zhang, R. Sankar, G. Q. Chang, Z. J. Yuan, C. C. Lee, S. M. Huang, H. Zheng, J. Ma, D. S. Sanchez, B. K. Wang, A. Bansil, F. C. Chou, P. P. Shibayev, H. Lin, S. Jia and M. Z. Hasan, *Discovery of a Weyl fermion semimetal and topological Fermi arcs.* Science **349**, 613 (2015).

[35] N. Xu, H. M. Weng, B. Q. Lv, C. Matt, J. Park, F. Bisti, V. N. Strocov, E. Pomjakushina, K. Conder, N. C. Plumb, M. Radovic, G. Autès, O.V. Yazyev, Z. Fang, X. Dai, G. Aeppli, T. Qian, J. Mesot, H. Ding and M. Shi, *Observation of Weyl nodes and Fermi arcs in tantalum phosphide.* Nat. Commun. **7**, 11006 (2016).

[36] L. X. Yang, Z. K. Liu, Y. Sun, H. Peng, H. F. Yang, T. Zhang, B. Zhou, Y. Zhang, Y. F. Guo, M. Rahn, D. Prabhakaran, Z. Hussain, S. K. Mo, C. Felser, B. Yan and Y. L. Chen, *Weyl semimetal phase in the non-centrosymmetric compound TaAs.* Nat. Phys. **11**, 728 (2015).

[37] N. Xu, G. Autès, C. E. Matt, B. Q. Lv, M. Y. Yao, F. Bisti, V. N. Strocov, D. Gawryluk, E. Pomjakushina, K. Conder, N. C. Plumb, M. Radovic, T. Qian, O. V. Yazyev, J. Mesot, H. Ding and M. Shi, *Distinct Evolutions of Weyl Fermion Quasiparticles and Fermi Arcs with Bulk Band Topology in Weyl Semimetals.* Phys. Rev. Lett. **118,** 106406 (2017).

[38] S. Y. Xu, Y. Xia, L. A. Wray, S. Jia, F. Meier, J. H. Dil, J. Osterwalder, B. Slomski, A. Bansil, H. Lin, R. J. Cava and M. Z. Hasan, *Topological Phase Transition and Texture Inversion in a Tunable Topological Insulator.* Science **332**, 560 (2011).

[39] N. Xu, Y. T. Qian, Q. S. Wu, G. Autès, C. E. Matt, B. Q. Lv, M. Y. Yao, V. N. Strocov, E. Pomjakushina, K. Conder, N. C. Plumb, M. Radovic, O. V. Yazyev, T. Qian, H. Ding, J. Mesot and M. Shi, *Trivial topological phase of CaAgP and the topological nodal-line transition in CaAg($P_{1-x}As_x$).* Phys. Rev. B **97**, 161111(R) (2018)

[40] N. Xu, Z. W. Wang, A. Magrez, P. Bugnon, H. Berger, C. E. Matt, V. N. Strocov, N. C. Plumb, M. Radovic, E. Pomjakushina, K. Conder, J. H. Dil, J. Mesot, R. Yu, H. Ding and M. Shi, *Evidence of a Coulomb-Interaction-Induced Lifshitz Transition and Robust Hybrid Weyl Semimetal in $T_d$-$MoTe_2$.* Phys. Rev. Lett. **121**, 136401 (2018).

[41] S. Li, Y Liu, S. S. Wang, Z. M. Yu, S. Guan, X. L. Sheng, Y. G. Yao and S. Y. A. Yang, *Nonsymmorphic-symmetry-protected hourglass dirac loop, nodal line, and dirac point in bulk and monolayer $X_3SiTe_6$ (X = Ta, Nb).* Phys. Rev. B **97**, 045131 (2018).

[42] T. Sato, Z. W. Wang, K. Nakayama, S. Souma, D. Takane, Y. Nakata, H. Iwasawa, C. Cacho, T. Kim, T. Takahashi and Y. Ando, *Observation of band crossings protected by nonsymmorphic symmetry in the layered ternary telluride $Ta_3SiTe_6$.* Phys. Rev. B **98**, 121111(R) (2018).

[43] L. L. An, H. W. Zhang, J. Hu, X. D. Zhu, W. S. Gao, J. L. Zhang, C. Y Xi, W. Ning, Z. Q Mao and M. L Tian, *Magnetoresistance and Shubnikov–de Haas oscillations in layered $Nb_3SiTe_6$ thin flakes.* Phys. Rev. B **97**, 235133 (2018).

[44] M. Naveed, F. C. Fei, H. J. Bu, X. Y. Bo, S. A. Shah, B. Chen, Y. Zhang, Q. Q. Liu, B. Y. Wei, S. Zhang, J. W. Guo, C. Y. Xi, A. Rahman, Z. M. Zhang, M. H. Zhang, X. G. Wan and F. Q. Song, *Magneto-transport and Shubnikov-de Haas oscillations in the layered ternary telluride topological semimetal candidate*

*Ta₃SiTe₆.* Appl. Phys. Lett. **116**, 092402 (2020).

[45] J. Hu, X. Liu, C. L. Yue, J. Y. Liu, H. W. Zhu, J. B. He, J. Wei1, Z. Q. Mao, L. Yu. Antipina, Z. I. Popov, P. B. Sorokin, T. J. Liu, P. W. Adams, S. M. A. Radmanesh, L. Spinu, H. Ji and D. Natelson, *Enhanced electron coherence in atomically thin Nb₃SiTe₆.* Nat. Phys. **11**, 471 (2015).

[46] T. Y. Yang, Q. Wan, D. Y. Yan, Z. Zhu, Z. W. Wang, C. Peng, Y. B. Huang, R. Yu, J. Hu, Z. Q. Mao, S. Li, S. Y. A. Yang, H. Zheng, J. F. Jia, Y. G. Shi and N. Xu, *Directional massless Dirac fermions in a layered van der Waals material with one-dimensional long-range order.* Nat. Mater. **19**, 27 (2020).

[47] V.N. Strocov, X. Wang, M. Shi, M. Kobayashi, J. Krempasky, C. Hess, T. Schmitt & L. Patthey, *Soft-X-ray ARPES facility at the ADRESS beamline of the SLS: Concepts, technical realisation and scientific applications*, J. Synchrotron Rad. **21**, 32 (2014).

[48] J. P. Perdew, K. Burke, and M. Ernzerhof, *Generalized gradient approximation made simple.* Phys. Rev. Lett. **77**, 3865 (1996).

[49] G. Kresse and J. Hafner, *Ab initio molecular-dynamics simulation of the liquid-metal–amorphous-semiconductor transition in germanium.* Phys. Rev. B **49**, 14251 (1994).

[50] G. Kresse and J. Furthmüller, *Efficient iterative schemes for ab initio total-energy calculations using a plane-wave basis set.* Phys. Rev. B **54**, 11169 (1996).

[51] P. E. Blöchl, *Projector augmented-wave method.* Phys. Rev. B **50**, 17953 (1994).

[52] M. Dion, H. Rydberg, E. Schröder, D. C. Langreth, and B. I. Lundqvist, *Van der Waals density functional for general geometries.* Phys. Rev. Lett. **92**, 246401 (2004).

[53] N. Marzari and D. Vanderbilt, *Maximally localized generalized Wannier functions for composite energy bands.* Phys. Rev. B **56**, 12847 (1997).

[54] I. Souza, N. Marzari, and D. Vanderbilt, *Maximally localized Wannier functions for entangled energy bands.* Phys. Rev. B **65**, 035109 (2001).

[55] M. P. López Sancho, J. M. López Sancho, and J. Rubio, *Quick iterative scheme for the calculation of transfer matrices: application to Mo (100).* J. Phys. F **14**, 1205 (1984)

[56] M. P. López Sancho, J. M. López Sancho, and J. Rubio, *Highly convergent schemes for the calculation of bulk and surface Green functions.* J. Phys. F **15**, 851 (1985).

[57] Q. S. Wu, S. N. Zhang, H.-F. Song, M. Troyer, and A. A. Soluyanov, *WannierTools: An open-source software package for novel topological materials.* Comput. Phys. Commun. **224**, 405 (2018).

[58] V. N. Strocov, M. Shi, M. Kobayashi, C. Monney, X. Wang, J. Krempasky, T. Schmitt, L. Patthey, H. Berger and P. Blaha, *Three-Dimensional Electron Realm in $VSe_2$ by Soft-X-Ray Photoelectron Spectroscopy: Origin of Charge-Density Waves*, Phys. Rev. Lett. **109**, 086401 (2012)

[59] D. Pescia, A. R. Law, M. T. Johnson and H. P. Hughes, *Determination of observable conduction band symmetry in angle-resolved electron spectroscopies: Non-symmorphic space groups.* Solid State Commun. **56,** 809 (1985).

[60] Th. Finteis, M. Hengsberger, Th. Straub, K. Fauth, R. Claessen, P. Auer, P. Steiner, and S. Hüfner, *Occupied and unoccupied electronic band structure of $WSe_2$.* Phys. Rev. B **55**, 10400 (1997).

[61] G. Landolt, S. V. Eremeev, O. E. Tereshchenko, S. Muff, B. Slomski, K. A. Kokh, M. Kobayashi, T. Schmitt, V. N Strocov, J. Osterwalder, E. V. Chulkov and J. H. Dil, *Bulk and surface Rashba splitting in single termination BiTeCl.* New J. Phys. **15**, 085022 (2013).

[62] T. Bzdušek, Q. S. Wu, A. Rüegg, M. Sigrist, and A. A. Soluyanov, *Nodal-chain metals.* Nature **538**, 75 (2016).

[63] X. W. Zhang, Q. H. Liu, J. W. Luo, A. J. Freeman, A. Zunger, *Hidden spin polarization in inversion-symmetric bulk crystals.* Nature Phys. **10**, 387–393 (2014).

[64] J. M. Riley, F. Mazzola, M. Dendzik, M. Michiardi, T. Takayama, L. Bawden, C. Granerød, M. Leandersson, T. Balasubramanian, M. Hoesch, T. K. Kim, H. Takagi, W. Meevasana, Ph. Hofmann, M. S. Bahramy, J. W. Wells, P. D. C. King, Nature Phys. **10**, 835–839 (2014).

[65] Z. Zhu, S. Li, M. Yang, X. A. Nie, H. K. Xu, X. Yang, D. D. Guan, S. Y. Wang, Y. Y. Li, C. H. Liu, Z. Q. Mao, N. Xu, Y. G. Yao, S. Y. A. Yang, Y. G. Shi, H. Zheng and J. F. Jia, *A tunable and unidirectional one-dimensional electronic system $Nb_{2n+1}SiNTe_{4n+2}$.* npj Quantum Materials **5**, 35 (2020).

**Figures**

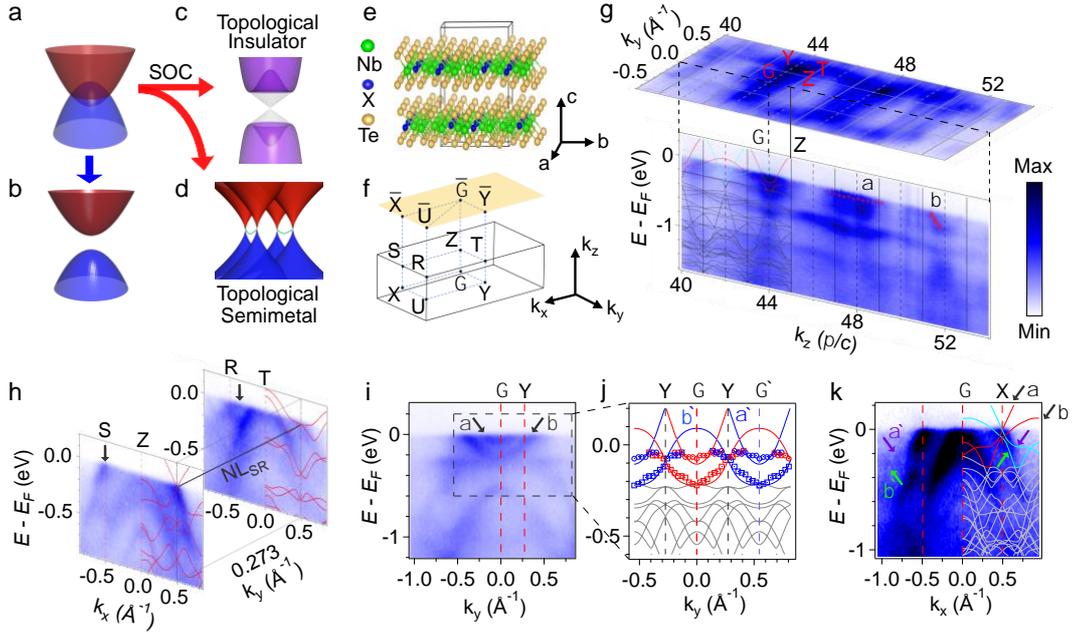

**Figure 1. (a-b)** Band structures with and without band inversion, respectively, when spin orbit coupling (SOC) is not taken into account. **(c-d)** SOC induced topological insulator and topological semimetal, respectively. **(e)** Crystal structure of $Nb_3XTe_6$. **(f)** Bulk and surface Brillouin zone with the high-symmetry points indicated. **(g)** Bulk band structure of $Nb_3SiTe_6$. The red symbols are extracted band dispersions. **(h)** Photoemission intensity plot along the Z-S and T-R direction. The black line represents Dirac nodal line from calculations. **(i,j)** Photoemission intensity plot along the Γ-Y direction. The red symbols indicate the extracted band structure and the blue ones are same as the red symbols but shifted by $2\pi/b$. **(k)** Photoemission intensity plot along the Γ-X direction. The calculations without SOC are appended for a direct comparison in **(g)**, **(h)**, **(j)** and **(k)**. The red and blue lines in calculations in **(g)**, **(j)** and **(k)** represent bands near $E_F$ with and without observable spectra weight in ARPES measurements, respectively.

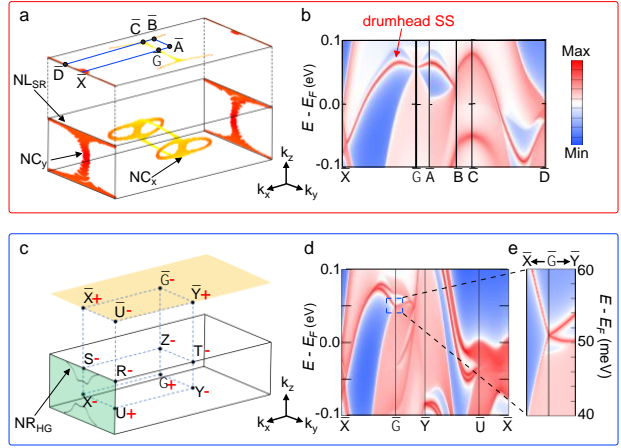

**Figure 2. (a)** The bulk nodes and their surface projections in Nb$_3$SiTe$_6$ from calculations without SOC. **(b)** The calculated spectra weight on the (001) surface, along the momentum path passing through the projections of nodes as indicated in **(a)**. **(c)** The calculated parity of time-reversal invariant momentum points in the bulk and surface Brillouin zones, as SOC is taken into account. **(d)** The spectra weight on the (001) surface derived from calculations with SOC. **(e)** A zoomed in area in **(d)**.

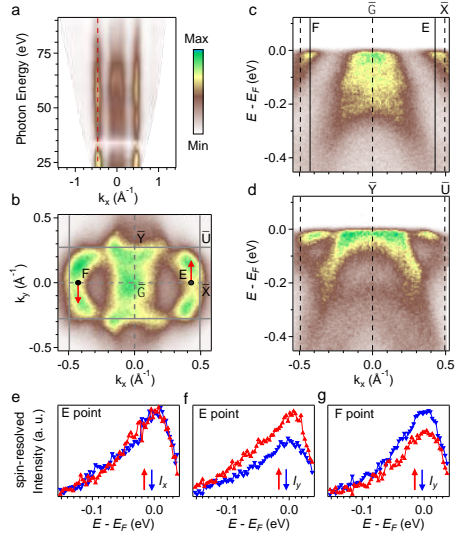

**Figure 3. (a)** Photon energy dependent ARPES results at $E_F$ of $Nb_3GeTe_6$ along the $\bar{\Gamma}$-$\bar{X}$ direction. **(b)** Fermi surface mapping of $Nb_3GeTe_6$. **(c,d)** Photoemission intensity plots along the $\bar{\Gamma}$-$\bar{X}$ and $\bar{Y}$-$\bar{U}$ directions, respectively. **(e)** Spin-polarized EDC along the x direction ($I^{\uparrow\downarrow}_x$) measured at the E point as indicated in **(b)** and **(c)**. The red and blue curves represent the positive and negative directions, respectively. **(f)** Same as **(e)** but for the spin polarization along the y direction ($I^{\uparrow\downarrow}_y$). **(g)** Same as **(f)**, but measured at the F point in **(b)** and **(c)**. **(b)-(g)** are measured on $Nb_3GeTe_6$ with 21.2 eV light.

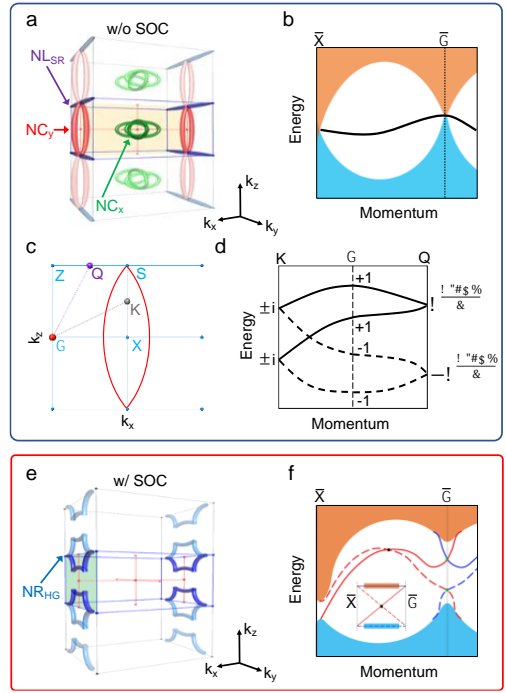

**Figure 4. (a-b)** Schematic drawings of Dirac nodal line/chains and topological surface states, respectively. **(c)** Dirac nodal chain $NC_y$ in the $k_y = 0$ plane. **(d)** Band structure without SOC along the K-Γ-Q direction, with the eigenvalues of $\widetilde{M_y}$ labeled. **(e)** Hourglass nodal ring as SOC is included. **(f)** Topological surface states as inherited signature of weak TI phase.